\documentclass[12pt]{article}
\usepackage{amsfonts}
\usepackage[utf8]{inputenc}
\usepackage{bm}
\usepackage{amsmath}
\usepackage{epsfig}
\usepackage{enumitem}

\setlist[enumerate]{%
wide =0.5\parindent,
listparindent=0pt%
}

\newcommand{\bmat}{\left(\begin{array}}
\newcommand{\emat}{\end{array}\right)}

\def\gtrsim{\mathrel{\raise.3ex\hbox{$>$\kern-.75em\lower1ex\hbox{$\sim$}}}}

\def\ov{\overline}
\def\un{\underline}

\def\-{\hphantom{-}}
\def\ov{\overline}
\def\s2{\frac{1}{\sqrt2}}

\def\mg{m_{3/2}}
\def\mg2{m^2_{3/2}}

\def\Dsl{\,\raise.15ex\hbox{/}\mkern-13.5mu D} 

\def\be{\begin{equation}}
\def\ee{\end{equation}}
\def\bea{\begin{eqnarray}}
\def\eea{\end{eqnarray}}

\newcommand{\nn}{\nonumber}

\begin{document}


\pagestyle{plain}

\makeatletter
\@addtoreset{equation}{section}
\makeatother
\renewcommand{\theequation}{\thesection.\arabic{equation}}
\pagestyle{empty}
\begin{center}
\ \

\vskip .5cm

\LARGE{\LARGE\bf On the phase space in Double Field Theory \\[10mm]}
\vskip 0.3cm

\large{Eric Lescano$^\dag$ and Nahuel Mir\'on-Granese$^*$
 \\[6mm]}

{\small  $^\dag$ Instituto de Astronom\'ia y F\'isica del Espacio (IAFE-CONICET-UBA)\\ [.01 cm]}
{\small\it Ciudad Universitaria, Pabell\'on IAFE, 1428 Buenos Aires, Argentina\\ [.3 cm]}
{\small  $^*$ Departamento de F\'isica, FCEyN, Universidad de Buenos Aires (UBA) \\ [.01 cm]}
{\small\it Ciudad Universitaria, Pabell\'on 1, 1428 Buenos Aires, Argentina\\ [.5 cm]}

{\small \verb"elescano@iafe.uba.ar, nahuelmg@df.uba.ar"}\\[1cm]

\small{\bf Abstract} \\[0.5cm]\end{center}
 
We present a model of (double) kinetic theory which paves the way to describe matter in a Double Field Theory background. Generalized diffeomorphisms acting on double phase space tensors are introduced. The generalized covariant derivative is replaced by a generalized Liouville operator as it happens in relativistic kinetic theory. 
The section condition is consistently extended and the closure of the generalized transformations is still given by the C-bracket. In this context we propose a generalized Boltzmann equation and compute the moments of the latter, obtaining an expression for the generalized energy- momentum tensor and its conservation law.        

\newpage

\setcounter{page}{1}
\pagestyle{plain}
\renewcommand{\thefootnote}{\arabic{footnote}}
\setcounter{footnote}{0}

\tableofcontents
\newpage

\section{Introduction}

The Einstein's field equations,
\bea
G_{\mu \nu} = T_{\mu \nu}
\label{Einsteinforever}
\eea
are fundamental relations which describe the dynamics of matter coupled to gravity in a Riemannian $D$-dimensional background ($\mu=0,\dots,D-1$). The LHS of the equation is given by the Einstein tensor, a divergenceless and symmetric tensor that depends on the geometric properties of the $D$-dimensional space-time and the RHS is related to the matter and energy content of the system.

Kinetic theory is the usual way to describe matter from microscopic principles. In this scheme the energy-momentum tensor is the second moment of the one-particle distribution function $f=f[x,p]$ \cite{Kinetic, Rezzolla, Cercignani},
\bea
T^{\mu\nu}=\int p^{\mu}p^{\nu} f \sqrt{g}\,d^Dp \, , \label{tmunuintroduction}
\eea
with $p^\mu$ the momentum and $g$ the determinant of the metric tensor. The evolution of $f$ is given by the relativistic Boltzmann equation
\bea
p^\mu D_\mu f=C[f] \, ,
\label{Boltzmannintro}
\eea
where $C[f]$ is the collision term and $D_{\mu}$ is the Liouville operator defined as
\bea
D_{\mu} = \nabla_{\mu} - \Gamma_{\mu \nu}^{\sigma} p^{\nu} \frac{\partial}{\partial p^{\sigma}} \, .
\eea
In the previous expression $\nabla_{\mu}$ is the covariant derivative of the space-time using the Levi-Civita connection $\Gamma_{\mu \nu}^{\sigma}$. For applications and extensions of the formalism see \cite{calzettahu}.

The Boltzmann equation (\ref{Boltzmannintro}) describes the evolution of the number of particles in a given volume of the $2D$-dimensional phase space, which in this context is defined as follows: for each point $x$ on the $D$-manifold $M$, a momentum space $\mathbb{P}_{x}$ attached to $x$ is introduced. Then the phase space is the collection $(x,\mathbb{P}_x)$ which defines a fiber bundle \cite{fiberbundle}. On account of this the RHS of the Einstein equation also holds a rich geometric structure that is part of the relativistic kinetic theory. For instance, the Liouville operator $D_{\mu}$ can be understood as a covariant derivative on the phase-space. In fact the infinitesimal diffeomorphisms for a generic phase-space tensor $v_{\mu}{}^{\nu}(x,p)$ receive an extra contribution of the form \cite{Sarbach},   
\bea
\delta_{\xi} v_{\mu}{}^{\nu} = L_{\xi} v_{\mu}{}^{\nu} + p^{\rho} \frac{\partial \xi^{\sigma}(x)}{\partial x^{\rho}} \frac{\partial v_{\mu}{}^{\nu}}{\partial p^{\sigma}} \, ,
\label{diffintro}
\eea
where $L_{\xi}$ is the Lie derivative acting on tensors defined as
\bea
L_{\xi} v_{\mu}{}^{\nu} = \xi^{\sigma} \frac{\partial v_{\mu}{}^{\nu}}{\partial x^{\sigma}} + \frac{\partial \xi^{\rho}}{\partial x^{\mu}} v_{\rho}{}^{\nu} - \frac{\partial \xi^{\nu}}{\partial x^{\rho}} v_{\mu}{}^{\rho} + \omega \frac{\partial \xi^{\sigma}}{\partial x^{\sigma}} v_{\mu}{}^{\nu} \, , 
\eea
where $\omega$ is a weight constant and $\xi_{\mu}=\xi_{\mu}(x)$ an infinitesimal parameter. The closure of (\ref{diffintro}) is given by the Lie bracket,
\bea
\xi^{\mu}_{12}(x) = \xi^{\rho}_{1} \frac{\partial \xi^{\mu}_{2}}{\partial x^{\rho}} - (1 \leftrightarrow 2)  \, .
\eea

On the other hand the energy-momentum tensor (\ref{tmunuintroduction}) is a symmetric and divergenceless tensor related with the variation of the matter action $S_{m}$ with respect to the (inverse) metric tensor as
\bea
T_{\mu \nu} = \frac{-2}{\sqrt g}\frac{\delta S_m
}{\delta g^{\mu \nu}} \, .\label{tmunuGR}
\eea
Then, (\ref{Einsteinforever}) is the equation of motion of the metric tensor when we couple matter to a Riemannian background.

In this work we present a model of kinetic theory which paves the way to describe matter in a Double Field Theory (DFT) background. DFT \cite{Siegel,DFT,DFTkorea} is a generalization of Riemannian geometry which is manifestly invariant under $O(D,D)$. The previous group is closely related with an exact symmetry of String Theory \cite{Ashoke}. However since the dimension of the fundamental representation of $O(D,D)$ is $2D$, the ordinary space-time must be doubled to accomplish $O(D,D)$ as a global symmetry of the theory \footnote{Check \cite{ReviewDFT} for reviews.}. The generalized coordinates of the double space $X^{M}=(x^{\mu},\tilde{x}_{\mu})$ are in the fundamental representation of $O(D,D)$, where $\tilde{x}_{\mu}$ is the extra set of coordinates and $M=0,\dots,2D-1$. Derivatives in the double space are constrained by the section condition (or strong constraint),
\be
\partial_{M} (\partial^{M} \star) = (\partial_{M} \star) (\partial^{M} \star) = 0 \, ,
\ee
where $\star$ means a product of arbitrary generalized fields. These constraints effectively removes the dependence on $\tilde{x}_{\mu}$.

The invariant metric of $O(D,D)$ is
\bea
{\eta}_{{M N}}  = \left(\begin{matrix}0&\delta_\mu^\nu\\ 
\delta^\mu_\nu&0 \end{matrix}\right)\, .  \label{etaintro}
\eea
This metric raises and lowers the indices $M,N,\dots$ and is left invariant under generalized diffeomorphisms, generated infinitesimally  by $\xi^{M}$  through the generalized Lie derivative, defined as
\bea
{\cal L}_\xi V_M(X) = \xi^{N} \partial_N V_M(X) + (\partial_M \xi^N - \partial^N \xi_{M}) V_N(X) + \omega (\partial_{N} \xi^{N}) V_{M}(X) \, ,
\label{glieintro}
\eea 
where $V_M(X)$ is an arbitrary (double) vector and $\omega$ is a weight constant. The closure of the generalized diffeomorphisms, 
\bea
\Big[\delta_{\xi_1},\delta_{\xi_2} \Big] V^{M}(X) = \delta_{\xi_{21}} V^{M}(X) 
\eea
is given by the C-bracket,
\bea
\xi^{M}_{12}(X) = \xi^{P}_{1} \frac{\partial \xi^{M}_{2}}{\partial X^{P}} - \frac12 \xi^{P}_{1} \frac{\partial \xi_{2P}}{\partial X_{M}} - (1 \leftrightarrow 2)  \, .
\eea 

The background field content of DFT consists of a dynamical metric ${\cal H}_{MN}(X)={\cal H}_{MN}$, called the generalized metric and a scalar $d(X)=d$, called the generalized dilaton. Both ${\cal H}_{M N}$ and $\eta_{M N}$ can be decompose in terms of the DFT projectors namely
\bea
P_{MN} = \frac{1}{2}\left(\eta_{MN} - {\cal H}_{MN}\right)  \ \ {\rm and} \ \
\ov{P}_{MN} = \frac{1}{2}\left(\eta_{MN} + {\cal H}_{MN}\right)\ ,
\eea
which satisfy the following properties 
\bea
&{\overline{P}}_{{M Q}} {\overline{P}}^{ Q}{}_{ N}={\overline{P}}_{{M N}}\, , &\quad {P}_{{M Q}} {P}^{Q}{}_{ N}={P}_{{M N}}, \nn\\
&{P}_{{M  Q}}{\overline{P}}^{Q}{}_{ N} = {\overline{P}}_{ {M Q}}  {P}^{ Q}{}_{ N} = 0\, ,  &\quad {\overline{P}}_{{MN}} + {P}_{{M N}} = \eta_{{M N}}\,.
\eea
Using the previous projectors, an arbitrary vector $V_{M}$  can be decomposed as
\bea
V_{M} = V_{\underline{M}} + V_{\overline M} = P_{M}{}^{N} V_{N} + \bar{P}_{M}{}^{N} V_{N} \, .
\eea
The partial derivative of a vector in the double space does not transform like a vector. The generalized covariant derivative is defined as
\bea
\nabla_{M} V_{N} = \partial_{M} V_{N} - \Gamma_{M N}{}^{P} V_{P} \, ,
\label{covariantintro}
\eea
and demanding compatibility with the invariant group, with the generalized metric and demanding that the generalized torsion vanishes is not enough to completely determine the generalized affine connection. However, the generalized Ricci tensor ${\cal R}_{MN}[{\cal H}, d]$ and the generalized Ricci scalar ${\cal R}[{\cal H}, d]$ are fully determined. In consequence the DFT action principle is defined in the following way,
\bea
\int d^{2D}X e^{-2d} \Bigg(\frac12\, {\cal R}[{\cal H},d] + {\cal L}_{m}[{\cal H},d,\Phi] \Bigg) 
\label{actionDFTintro}
\eea
where  ${\cal L}_{m}[{\cal H},d,\Phi]$ is the matter Lagrangian that may depend on extra fields, represented in (\ref{actionDFTintro}) as $\Phi$. Following a variational principle for this action with respect to generalized diffeomorphims \cite{Currents}, namely 
\bea
\delta_{\xi} ({\cal S}_{\rm 0}+{\cal  S}_m)=0 \, ,
\eea
it is possible to define an Einstein-type equation \cite{Park} of the form \bea
{\cal G}_{MN}={\cal T}_{MN}
\label{Einsteinintro}
\eea
where the generalized symmetric Einstein tensor results
\bea
{\cal G}_{MN} = -2({\cal R}_{\overline M \underline N} + {\cal R}_{\underline M \overline N}) - \frac12 {\cal H}_{M N} {\cal R} \, ,
\eea
and the definition of the generalized symmetric energy-momentum tensor is
\bea
{\cal T}_{MN} = {\cal H}_{MN}\left({\cal L}_m-\frac12\frac{\delta {\cal L}_m}{\delta d}\right)-2\,\Big[\overline{P}_{MK} P_{NL}+\overline{P}_{NK} P_{ML}\Big]\left(\frac{\delta {\cal L}_m}{\delta {P}_{KL}}-\frac{\delta {\cal L}_m}{\delta \overline{P}_{KL}}\right)\label{TMNlagrangianintro}\,.
\eea
The differential Bianchi identities in the double geometry reads \cite{FrameDFT, BianchiDFT},
\bea
\nabla_{\underline P}{\cal R} - 4 \nabla^{\overline M} {\cal R}_{\underline P \overline M} & = & 0 \, , \nn \\ \nabla_{\overline P}{\cal R} + 4 \nabla^{\underline M} {\cal R}_{\underline M \overline P } & = & 0 \, ,
\label{BianchiIntro}
\eea
and both of them provide $\nabla^{M}{\cal G}_{M N}=0$ off-shell. Similarly to what happens in General Relativity (GR) the generalized symmetric energy-momentum tensor is divergenless on-shell. The main goal of this work is to elaborate on the double phase space of DFT and to define an $O(D,D)$ invariant kinetic theory. In particular we study the relation between the symmetric energy-momentum tensor (\ref{TMNlagrangianintro}) and the second moment of the generalized distribution function, coming from the generalized Boltzmann equation.

\subsection{Main results}

We start defining a generalized momentum space $\mathbb{P}_X$ attached to the double space. The double phase space is the collection  
\bea
\Big\{X^{M},{\cal P}^{M} \Big\} \, ,
\eea
where ${\cal P}^{M}$ is an extra coordinate that is also an $O(D,D)$ vector. The infinitesimal generalized diffeomorphisms acting on a double phase space vector $V^Q=V^{Q}(X,{\cal P})$ are defined as
\bea
\delta_{\xi} V^{Q} = {\cal L}_{\xi} V^{Q} + {\cal P}^{N} \frac{\partial \xi^{M}}{\partial X^{N}} \frac{\partial V^Q}{\partial {\cal P}^{M}} - {\cal P}^{N} \frac{\partial \xi_{N}}{\partial X_{M}} \frac{\partial V^Q}{\partial {\cal P}^{M}}  \, ,
\label{transintro}
\eea
where ${\cal L}_{\xi}$ is the generalized Lie derivative and $\xi_M=\xi_M(X)$ is an infinitesimal parameter. We demand a section condition for the momentum derivatives,
\bea
(\frac{\partial}{\partial {\cal P}^{M}} \star) (\frac{\partial}{\partial {\cal P}_{M}} \star) = \frac{\partial}{\partial {\cal P}^{M}} (\frac{\partial}{\partial {\cal P}_{M}} \star) = 0 \,   
\eea
and for the mixed derivatives,
\bea
(\frac{\partial}{\partial X^{M}} \star) (\frac{\partial}{\partial {\cal P}_{M}} \star) = \frac{\partial}{\partial X^{M}} (\frac{\partial}{\partial {\cal P}_{M}} \star) = 0 \, . \,   
\eea
Imposing the previous constraints, the last term of (\ref{transintro}) vanishes. We solved the strong constraint of the double phase space with the following solution
\bea
\frac{\partial}{\partial {\cal P}_{\mu}} = 0 \, , 
\eea
which is enough to recover the usual phase space diffeomorphisms from the generalized ones and then ${\cal P}^{\mu}=p^{\mu}$ is the momentum. The closure of (\ref{transintro}), 
\bea
\Big[\delta_{\xi_1},\delta_{\xi_2} \Big] V^{M}(X,{\cal P}) = \delta_{\xi_{21}} V^{M}(X,{\cal P}) 
\eea
is given by the C-bracket,
\bea
\xi^{M}_{12}(X) = \xi^{P}_{1} \frac{\partial \xi^{M}_{2}}{\partial X^{P}} - \frac12 \xi^{P}_{1} \frac{\partial \xi_{2P}}{\partial X^{M}} - (1 \leftrightarrow 2)  \, .
\eea

The covariant derivative of the double phase space is given by the generalized Liouville operator $D_{M}$ defined as 
\bea
D_{M} = \nabla_{M} - \Gamma_{M N}{}^{Q} {\cal P}^{N} \frac{\partial}{\partial {\cal P}^{Q}} \, .
\eea
We define a generalized Boltzmann equation considering a generalized distribution function $F=F\big( X,{\cal P} \big)$ in the following way, 
\bea
{\cal P}^{M} {\cal D}_{M} F = {\cal C}[F] \, ,
\label{BoltzmannDFTintro}
\eea
where  ${\cal C}[F]$ is a generalized collision term and the operator ${\cal D}_M$ contains a generalized dilaton dependence,
\bea
{\cal D}_{M} = D_{M} - {\cal U}_{M} \, ,
\eea
with ${\cal U}_{M}= {\cal U}_{M}(X) = 2\,\partial_M d.$ In terms of the general relativistic Boltzmann equation, the ${\cal U}_M$ contribution can be interpreted as a generalized force term that appears since the integral measure of DFT is a function of the generalized dilaton. We assume the existence of an equilibrium state such that
\bea
{\cal C}[F_{eq}]=0 \, .
\eea
The integration of the product of (\ref{BoltzmannDFTintro}) and a generic phase space object $\Psi^{M}$ leads to  
\bea
\nabla_{N}\Bigg[\int \Psi^{M} {\cal P}^{N} F_{eq}\, e^{-2d} d^{2D}{\cal P} \Bigg] - \int F_{eq}\, {\cal P}^{N} {D}_{N} \Psi^{M} e^{-2d} d^{2D}{\cal P}  = 0 \, .
\label{lawDFTintro}
\eea
If we set $\Psi^{M}$ to a constant, we find the generalized current
\bea
{\cal N}^{M}(X)=\int {\cal P}^{M} F_{eq}\, e^{-2d} d^{2D}{\cal P}
\eea
with the conservation law $\nabla_{M} {\cal N}^{M} = 0$. Considering $\Psi^{M}={\cal P}^{M}$ in (\ref{lawDFTintro}) leads to the generalized energy-momentum tensor 
\bea
{\cal T}^{M N}(X) = \int {\cal P}^{M} {\cal P}^{N} F_{eq}\, e^{-2d} d^{2D}{\cal P} \, ,\label{emtintro}
\eea
and its conservation law reads $\nabla_{N}{\cal T}^{M N} = 0$.

Similarly to what occurs in GR, the present results show that the RHS of the generalized Einstein equation (\ref{Einsteinintro}) has the same properties, symmetric and divergenceless, than the generalized symmetric energy-momentum tensor coming from the double kinetic theory (\ref{emtintro}).

\subsection{Outline}
This work is organized as follows. In Section \ref{RKT} we review the main aspects of general relativistic kinetic theory and its Lagrangian and Hamiltonian formulation. We focus on the construction of the phase space and its symmetries. Special attention is paid to the Liouville operator, which is defined as the natural extension of the covariant derivative on the phase space. Then we summarize the construction of the Boltzmann equation and the conservation laws for the first and second moment of the distribution function.

Section \ref{DFT} is dedicated to the basic aspects of the geometry of DFT. We start defining generalized diffeomorphisms through a generalized Lie derivative and then we introduce a generalized affine connection that, unlike GR, is undetermined. After that we review the differential Bianchi identities, the generalized Einstein tensor and the  generalized energy-momentum tensor. 

In Section \ref{DKT} we elaborate on the kinetic theory in the double space. We start by giving a consistent deformation of the generalized diffeomorphisms with an appropriate section condition for the generalized momentum derivatives. The bracket of the previous transformations is the C-bracket and the generalized Liouville operator plays the role of the covariant derivative on the double space. Then we present a generalized Boltzmann equation and a generalized  transfer equation, and we obtain the conservation laws for the generalized current and the generalized energy-momentum tensor. In the last part of this Section we discuss applications. Here we focus in the generalized scalar field Lagrangian as a candidate of matter Lagrangian for a perfect fluid in the double space. 

In Section \ref{Outlook} we conclude the work and elaborate on some future directions. After that two appendices are provided. In \ref{Appendix} we set our conventions and in \ref{Clos} we explicitly show the closure of the generalized diffeomorphism transformations in the double phase space.

\section{Relativistic Kinetic Theory}
\label{RKT}

\subsection{Basics}

We start with a $D$-manifold $M$ with coordinates $x^{\mu}$, $\mu=0, \dots, D$, equipped with a metric tensor $g_{\mu \nu}$ and a Levi-Civita connection $\Gamma_{\mu \nu}^{\rho}$.
For each point $x$ on $M$ with coordinate $x^\mu$, we introduce its tangent space $\mathbb{P}_{x}$ whose vectors are the momenta $p^\mu$. In consequence the phase space is a collection $(x,\mathbb{P}_x)$ which defines a tangent bundle \cite{fiberbundle, Sarbach}.
From this point of view, the momentum can be considered independent of the position and thus
\bea
\frac{\partial p^{\nu}}{\partial x^{\mu}} = 0 \, .
\label{IndepGR}
\eea
This condition holds in an off-shell formulation of the general relativistic kinetic theory which is the scenario that we will deal with. Conversely the on-shell condition $p^\mu p_\mu=m^2$ spoils the independence between momenta itself and with position coordinates.

Let us observe that coordinate transformations on $M$, induce transformations in the fiber
and therefore the infinitesimal diffeomorphisms of a phase space scalar $v$ with constant weight  $\omega$ can be written as
\bea
\delta_{\xi} v = L_{\xi} v + p^{\rho} \frac{\partial \xi^{\sigma}(x)}{\partial x^{\rho}} \frac{\partial v}{\partial p^{\sigma}} \, ,
\label{ediffeos}
\eea
where $L_{\xi}$ is the usual Lie derivative defined as
\bea
L_{\xi} v = \xi^{\sigma} \frac{\partial v}{\partial x^{\sigma}} + \omega \frac{\partial \xi^{\sigma}}{\partial x^{\sigma}} v \, , 
\eea
with $\xi^{\mu}=\xi^{\mu}(x)$ an infinitesimal parameter that characterizes the transformation. We may extend (\ref{ediffeos}) to tensors by taking the usual Lie derivative acting on different tensor structures, {\it e.g}. for a (1,1) tensor we have
\bea
L_{\xi} v_{\mu}{}^{\nu} = \xi^{\sigma} \frac{\partial v_{\mu}{}^{\nu}}{\partial x^{\sigma}} + \frac{\partial \xi^{\rho}}{\partial x^{\mu}} v_{\rho}{}^{\nu} - \frac{\partial \xi^{\nu}}{\partial x^{\rho}} v_{\mu}{}^{\rho} + \omega \frac{\partial \xi^{\sigma}}{\partial x^{\sigma}} v_{\mu}{}^{\nu} \, .\label{infdiftensor}
\eea

It is straightforward to check the closure of the transformation (\ref{infdiftensor}),
\bea
\Big[\delta_{\xi_1},\delta_{\xi_2} \Big] v_{\mu}{}^{\nu} = \delta_{\xi_{21}} v_{\mu}{}^{\nu}
\eea
and show that the bracket is given by the Lie bracket, 
\bea
\xi^{\mu}_{12}(x) = \xi^{\rho}_{1} \frac{\partial \xi^{\mu}_{2}}{\partial x^{\rho}} - (1 \leftrightarrow 2)  \, .
\eea

Since we have tensors acting on the phase space we need to define a natural extension of the covariant derivative in the phase space, namely the Liouville operator $D_{\mu}$ (see {\it e.g.} Appendix A of \cite{Esteban}). Regarding that we have taken the collection $(x^\mu,p^\mu)$ to be the basis of the phase space, the Liouville operator for an arbitrary tensor reads
\bea
D_{\mu} A^{\rho\lambda}(x,p) = \nabla_{\mu} A^{\rho\lambda}(x,p) -  \Gamma^{\sigma}_{\mu \nu} p^{\nu}\frac{\partial A^{\rho\lambda}(x,p)}{\partial p^{\sigma}} \, ,\label{LiouvilleoperatorGR}
\eea
where $\nabla_\mu$ is the well-known covariant derivative. In particular it satisfies
\bea
D_{\mu} p^{\nu} = D_{\mu} p_{\nu} = 0 \, .
\eea

Finally the diffeomorphism invariant volume element of the phase space is the product of the coordinate and momentum invariant volume elements, namely
\bea
\sqrt g d^{d}{p}\,\sqrt g d^dx=g d^{d}{p} d^dx \, ,
\label{GRvolume}
\eea
with $g$ the determinant of the metric tensor.

\subsection{Lagrangian and Hamiltonian formulations}
\label{Hformulation}

In GR the minimum action principle for the trajectories of freely falling particles is 
\bea
S=\int ds\,,\label{actionGRfreelyfalling}
\eea
with $ds^2=g_{\mu\nu}(x)\,dx^\mu dx^\nu$. Since this action is invariant under both diffeomorphisms and reparametrizations, we may rewrite (\ref{actionGRfreelyfalling}) parametrized by the affine parameter $\lambda$ as
\bea
S=\int d\lambda\;L(x,dx/d\lambda,\lambda)\quad\quad{\rm with}\quad\quad L = \frac12\,g_{\mu\nu}(x)\frac{dx^\mu}{d\lambda}\frac{dx^\nu}{d\lambda}.\label{lagrangianforhamilton}
\eea
Naturally the equations of motion computed from (\ref{actionGRfreelyfalling}) are equivalent to those derived from (\ref{lagrangianforhamilton}). Therefore these equations are
\bea
\frac{d^2x^\mu}{d\lambda}+\Gamma^{\mu}_{\rho\sigma}\frac{dx^\rho}{d\lambda}\frac{dx^\sigma}{d\lambda}=0,\label{geodesicequationx}
\eea
where $\Gamma^{\mu}_{\rho\sigma}$ is the already mentioned Levi-Civita connection and $x^\mu$ and  $dx^\mu/d\lambda$ are the coordinates and its parameter-derivative respectively. This Lagrangian $L$ is a suitable choice to construct a manifestly covariant Hamiltonian formalism in which we define the canonical coordinates and conjugate momenta as $x^\mu$ and
\bea
p_\mu=\frac{\partial L}{\partial\left(dx^\mu/d\lambda\right)}=g_{\mu\nu}\frac{dx^\mu}{d\lambda}.
\eea
It turns out that the canonical variables $\{x^\mu,p^\mu\}$ define the phase space \cite{ wald, schutz}. In this scheme both coordinates and momenta are in an equal footing and must be varied independently. Further it is possible to write the Hamiltonian function as usual
\bea
H(x,p,\lambda)=p_\mu\frac{d x^\mu}{d\lambda}-L(x,d x /d\lambda,\lambda).\label{Hamiltonianexpression}
\eea

From (\ref{Hamiltonianexpression}) it is straightforward to show that the Hamilton's equations in covariant tensor form are
\bea
\frac{dx^\mu}{d\lambda}=\frac{\partial H}{\partial p_\mu}, \quad\quad\frac{dp_\mu}{d\lambda}=-\frac{\partial H}{\partial x^\mu}.\label{hamiltonequations}
\eea
The explicit form of the Hamiltonian function (\ref{Hamiltonianexpression}) is strictly based on the variational principle of the Lagrangian $L$. In this case we are dealing with trajectories of freely falling particles and the Hamiltonian function is proportional to the mass-shell condition,
\bea
H=\frac12\,g_{\mu\nu}(x)p^\mu p^\nu \, .
\label{ExplicitHamiltonianGR}
\eea
Equations (\ref{hamiltonequations}) are the same as (\ref{geodesicequationx}) regarding that $p^\mu=dx^\mu/d\lambda$.

\subsection{The relativistic Boltzmann equation}

The relativistic Boltzmann equation rules the evolution of the one-particle distribution function (1pdf) $f=f[x,p]$, which is a phase space scalar. In its simplest form this equation is \cite{Cercignani}
\bea
p^{\mu} D_{\mu} f = C[f] \, .\label{boltzmanneq}
\eea
The RHS of (\ref{boltzmanneq}) is the collision term which takes into account the non-gravitational interactions between particles. If an equilibrium state is achieved the 1pdf takes its equilibrium form $f=f_{\rm eq}$ and $C[f_{\rm eq}]=0$.

In this context we want to extract the geometric properties of the first and second moment of the distribution function in order to present the so-called transfer equations for the particle current and the energy-momentum tensor. We start by integrating the product of the relativistic Boltzmann equation and an arbitrary 1-index object of the phase space $\Psi^{\nu}[x,p]$, over the phase space, {\it i.e.},
\bea
\int \Psi^{\nu}(p^{\mu} \frac{\partial f}{\partial x^{\mu}} - \frac{\partial f}{\partial p^{\sigma}} \Gamma^{\sigma}_{\mu \rho} p^{\mu} p^{\rho})gd^dpd^dx = \int \Psi^{\nu} C[f] g d^dp d^dx \, .
\label{transfer}
\eea
When considering an equilibrium state the RHS of (\ref{transfer}) is vanishing as we have mentioned above. The LHS of (\ref{transfer}) can be simplified using the Leibniz rule, the equation
\bea
{p^{\mu} \frac{\partial g}{\partial x^{\mu}} - \frac{\partial}{{\partial p^{\sigma}}} (g \Gamma^{\sigma}_{\mu \nu} p^{\mu} p^{\nu}}) = 0 \, ,
\label{Liou}
\eea
which follows from (\ref{tracec}), and the divergence theorem 
\bea
\int  \frac{\partial }{\partial p^{\sigma}} (f \Psi^{\nu} \Gamma^{\sigma}_{\mu \rho} p^{\mu} p^{\rho}) gd^dpd^dx = 0 \,.
\eea
Thus we obtain the following conservation laws, independently of the integration on $\sqrt{g}d^dx$, 
\bea
\nabla_{\mu}\Big[\int \Psi^{\nu} p^{\mu} f \sqrt{g} d^dp \Big] - \int f p^{\mu} \nabla_{\mu} \Psi^{\nu}  \sqrt{g} d^dp + \int \frac{\partial \Psi^{\nu}}{\partial p^{\sigma}} \Gamma^{\rho}_{\mu \sigma} p^{\mu} p^{\sigma} f \sqrt{g} d^dp = 0 \, .
\label{law}
\eea
If we set the arbitrary function to a constant scalar, we have
\bea
\int p^{\mu} f[x,p] \sqrt{g} d^dp = N^{\mu}(x)
\eea
with the usual conservation law or transfer function for the particle current,
\be
\nabla_{\mu}N^{\mu} = 0 \, .
\label{c1}
\ee
If we instead take $\Psi^{\nu}=p^{\nu}$, since $p^{\nu}$ is in the kernel of $D_{\mu}$, the remaining terms in (\ref{law}) cancel out and we finally get the expression of the energy-momentum tensor 
\bea
\int p^{\mu} p^{\nu} f[x,p] \sqrt{g} d^dp = T^{\mu \nu}(x) \, ,
\eea
and its conservation law
\bea
\nabla_{\mu}T^{\mu \nu} = 0 \, 
.
\label{c2}
\eea

Interestingly enough, if we leave the equilibrium state, the conservation laws (\ref{c1}) and (\ref{c2}) still hold since the zeroth and first order moment of the collision term are vanishing for any $f$. This is so due to $p^{\nu}$ is a collisional conserved quantity or, in other words, a summational invariant \cite{Cercignani}. 

Before moving on let us briefly comment about the connection of the Boltzmann equation and the Hamiltonian formulation of the phase space. Every parameterized curve of the phase space has a tangent vector $\vec{L}$ described by
\bea
\vec L = \frac{dx^\mu}{d\lambda}\frac{\partial}{\partial x^\mu}+\frac{dp_\mu}{d\lambda}\frac{\partial}{\partial p_\mu}\label{tangentvector} \, .
\eea
This vector is related to the fundamental symplectic two-form \cite{ Sarbach, wald, schutz}
\bea
\bm \omega = \bm d p_\mu \wedge \bm{d} x^\mu\,,
\eea
which gives the symplectic structure to the phase space. The symplectic two-form fulfills the role of mapping vectors into one-forms. Particularly we obtain the Hamilton's equations by applying $\bm \omega$ to the tangent vector to the phase space trajectories, $\vec L$ (\ref{tangentvector}), and equating this result to the exterior derivative of the Hamiltonian function, namely
\bea
\bm \omega(\,\cdot\,,\vec L)&=&\bm d H(\,\cdot\,)\,,\label{symplecticequation}\\
\frac{dx^\mu}{d\lambda}\bm d p_\mu-\frac{d p_\mu}{d\lambda}\bm d x^\mu&=&\frac{\partial H}{\partial p_\mu}\bm d p_\mu+\frac{\partial H}{\partial x^\mu}\bm d x^\mu.
\label{Hamilton}
\eea

Finally there is an interesting property of the symplectic structure. Given $\bm \omega$ and $H$, the equation (\ref{symplecticequation}) may be understood as a relation which fixes the vector $\vec L$, also known as the Liouville vector. For this case, trajectories of freely falling particles, $H$ is set by (\ref{ExplicitHamiltonianGR}) thus we get
\bea
\vec L = g^{\mu\nu} p_\nu \frac{\partial}{\partial x^\mu}+g^{\rho \lambda}p_\lambda\Gamma^{\sigma}_{\mu\rho}\,p_\sigma \frac{\partial}{\partial p_\mu}\,.\label{Liouvillevector}
\eea
The expression above coincides with the Liouville operator for phase space scalars defined in (\ref{LiouvilleoperatorGR}) projected along the curve, {\it i.e.} $\vec L(\,\cdot\,)=p^\mu D_\mu (\,\cdot\,)$. From this point of view it is clear that the Boltzmann equation (\ref{boltzmanneq}) contains information about the geodesic trajectories following by the particles between collisions.

\section{Double Field Theory}
\label{DFT}

\subsection{Double space and generalized fields}

The geometry of DFT is based on a double space equipped with two metrics. On the one hand, we have the invariant metric of $O(D,D)$, $\eta_{M N}$, where $2D$ is the amount of dimensions of the theory. The indices $M,N,\dots$ are in the fundamental representation of $O(D,D)$ and are raised and lowered with $\eta^{M N}$ and $\eta_{M N}$ respectively. On the other hand we have the generalized metric ${\cal H}_{M N}$ that encodes the field content of the universal NS-NS sector of the low energy effective superstring theory, namely, a metric tensor $g_{\mu \nu}$, a Kalb Ramond field $b_{\mu \nu}=-b_{\nu \mu}$ and the dilaton $\phi$. The generalized metric is an element of $O(D,D)$ and therefore satisfies
\bea
{\cal H}_{M P} \eta^{P Q} {\cal H}_{Q N} = \eta_{M N} \, .
\label{Odd}
\eea
The main purpose of DFT is to define a theory manifestly invariant under $O(D,D)$, which is closely related with a symmetry of String Theory \cite{Ashoke}. In consequence all the DFT fields and parameters are $O(D,D)$ multiplets or group-invariant objects. Since the dimension of the fundamental representation of $O(D,D)$ is $2D$, the coordinates of DFT are $X^{M}=(x^{\mu},\tilde{x}_{\mu})$. The coordinates $\tilde{x}_{\mu}$ are known as the dual coordinates and are taken away imposing the strong constraint,
\be
\partial_{M} (\partial^{M} \star) = (\partial_{M} \star) (\partial^{M} \star) = 0 \, ,
\ee
where $\star$ means a product of arbitrary generalized fields. In this Section the notation for the derivatives is not ambiguous and $\partial_{M}=\frac{\partial}{\partial X^{M}}$. Using the previous constraints, the components of the fields and parameters of DFT depend only on $x^{\mu}$. The parametrization of the invariant metric is, 
\bea
{\eta}_{{M N}}  = \left(\begin{matrix}0&\delta_\nu^\mu\\ 
\delta^\nu_\mu&0 \end{matrix}\right)\, ,  \label{eta}
\eea
while the parametrization of the generalized metric is
\bea
{\cal H}_{{M N}}  = \left(\begin{matrix}g^{\mu \nu}&-g^{\mu \sigma}b_{\sigma \nu}\\ 
b_{\mu \sigma}g^{\sigma \nu}&g_{\mu \nu}-b_{\mu \sigma} g^{\sigma \rho} b_{\rho \nu} \end{matrix}\right)\, . 
\label{Gmetric}
\eea
It is straightforward to check that the previous parametrization satisfies (\ref{Odd}). 

In addition to the global $O(D,D)$ symmetry, DFT is invariant under generalized diffeomorphisms generated infinitesimally  by $\xi^{M}$  through the generalized Lie derivative, defined by 
\bea
{\cal L}_\xi V_M = \xi^{N} \partial_N V_M + (\partial_M \xi^N - \partial^N \xi_{M}) V_N + \omega (\partial_{N} \xi^{N})V_{M} \, ,
\label{glie}
\eea 
where $V_M$ is an arbitrary vector and $\omega$ is a weight constant. This expression is trivially extended to other tensors with different index structure. For example, the generalized metric is a generalized tensor with vanishing weight and $\eta_{M N}$ is trivially invariant.

The background field content of DFT consists of a generalized dilaton $d$ in addition to the generalized metric ${\cal H}^{MN}$. The former transforms as a tensor with weight $\omega=1$ and thus
\bea
\delta_{\xi} (e^{-2d}) = \partial_{P}(\xi^{P} e^{-2d}) \, , 
\eea
which means that $e^{-2d}$ transforms as a density and $e^{-2d}\,d^{2D}X$ defines the invariant volume of DFT. Similarly to GR, the partial derivative of a tensor does not transform as a tensor and therefore a covariant derivative must be included. We develope this issue in the next part of the work. 

\subsection{Generalized affine connection}
Having defined a generalized Lie derivative, it is natural to seek a covariant derivative. The latter is defined as
\bea
\nabla_{M} V_{N} = \partial_{M} V_{N} - \Gamma_{M N}{}^{P} V_{P} \, ,
\label{covderdft}
\eea
with trivial extension to tensors with more indices. Here we have introduced a generalized affine connection $\Gamma_{M N}{}^{P}$ whose transformation properties must compensate the failure of the partial derivative of a tensor to transform covariantly under
generalized diffeomorphisms. 

We can now demand some properties on the connection, namely:
\begin{itemize}
\item \underline{Compatibility with $\eta_{MN}$}:
\bea
\nabla_{M} \eta_{N P} = 0 \, ,
\eea
and then the generalized affine connection is antisymmetric in its last two indices, {\it i.e.}
\bea
\Gamma_{M N}{}^{Q} \eta_{Q P} = \Gamma_{M N P} = - \Gamma_{M P N} \, .
\eea
\item \underline{Compatibility with ${\cal H}_{MN}$}: 
\bea
\nabla_{M} {\cal H}_{N P} = 0 \, .
\eea
In order to discuss this item is convenient to define the $O(D,D)$ projectors,
\bea
P_{MN} = \frac{1}{2}\left(\eta_{MN} - {\cal H}_{MN}\right)  \ \ {\rm and} \ \
\ov{P}_{MN} = \frac{1}{2}\left(\eta_{MN} + {\cal H}_{MN}\right)\ ,
\eea
which satisfy the following properties 
\bea
&{\overline{P}}_{{M Q}} {\overline{P}}^{ Q}{}_{ N}={\overline{P}}_{{M N}}\, , &\quad {P}_{{M Q}} {P}^{Q}{}_{ N}={P}_{{M N}}, \nn\\
&{P}_{{M  Q}}{\overline{P}}^{Q}{}_{ N} = {\overline{P}}_{ {M Q}}  {P}^{ Q}{}_{ N} = 0\, ,  &\quad {\overline{P}}_{{MN}} + {P}_{{M N}} = \eta_{{M N}}\,.
\eea

The projections $\Gamma_{\overline{MNP}}$ and $\Gamma_{\underline{MNP}}$ remains undetermined after imposing $\nabla_{M}P_{NP}=0$ and $\nabla_{M} \bar{P}_{NP}=0$.

\item Partial integration in the presence of the generalized density $e^{-2d}$:
\bea
\int e^{-2d} V \nabla_{M} V^{M} d^{2D}X = - \int e^{-2d} V^{M} \nabla_{M} V d^{2D}X\label{partialintegrationcoordinates}
\eea
for arbitrary $V$ and $V^{M}$. The previous item forces 
\bea
\Gamma_{MN}{}^{M} = -2 \partial_{N} d \, .
\eea
\item Vanishing torsion:
\bea
\Gamma_{[MNP]} = \frac13 \,T_{MNP} = 0 \, .
\eea
Let us observe that the generalized torsion $T_{MNP}$ is antisymmetric in all its indices and transforms as a tensor (unlike $\Gamma_{[MN]P}$).   
\end{itemize}

As we have showed, while in GR demanding metric compatibility and vanishing torsion determines the connection completely (the affine connection turns to the Levi-Civita connection), in this approach of DFT these requirements turn out to leave undetermined components in the generalized version of the affine connection \cite{BianchiDFT}. At this point it is important to mention that there exists an equivalent scheme of DFT known as semi-covariant formalism in which the generalized connection is fully determined \cite{DFTkorea}. 

In the current case, the generalized Riemann tensor of the theory is undetermined but it is possible to take traces on it and obtain a generalized Ricci scalar ${\cal R}$ that is fully determined as a function of the generalized metric and the generalized dilaton,
\bea
{\cal R} = && \frac14 {\cal H}^{MN} \partial_{M}{\cal H}^{KL}\partial_{N}{\cal H}_{KL} - {\cal H}^{MN}\partial_{N}{\cal H}^{KL}\partial_{L}{\cal H}_{MK} + 8 {\cal H}^{MN} \partial_{M}\partial_{N}d \nn \\ && + 8 \partial_{M}{\cal H}^{MN} \partial_{N}d - 8 {\cal H}^{MN} \partial_{M}d \partial_{N}d - 2 \partial_{M} \partial_{N} {\cal H}^{MN} \, .
\label{scalarDFT}
\eea
In the next part of the work we discuss the differential Bianchi identities and the generalized energy-momentum tensor of DFT.

\subsection{Matter Lagrangian and the generalized energy-momentum tensor}
\label{Energy}
The action principle of DFT is
\bea
{\cal S} =\int d^{2D}X e^{-2d}\, {\cal L}\left[{\cal H},d,\Phi \right] = \int d^{2D}X e^{-2d} \Big(\frac12{\cal R}\left[{\cal H},d\right] +  {\cal L}_{m}\left[{\cal H},d,\Phi\right] \Big) \, ,
\label{actionDFT}
\eea
where ${\cal L}_{m}$ represents matter coupled to the background field content with matter fields collected in the notation $\Phi$. Using (\ref{scalarDFT}), parametrizing the generalized metric as showed in (\ref{Gmetric}), the generalized dilaton as
\bea
e^{-2d} = \sqrt{g} e^{-2\phi} \, ,
\eea
and imposing the strong constraint the DFT action reduces to the following action
\bea
S= \frac12 \int d^{D}x \sqrt{g} e^{-2\phi} \Bigg(R + 4 (\partial \phi)^2 - \frac{1}{12} H_{\mu \nu \rho} H^{\mu \nu \rho}  + 2 L_{m}[g,b,\phi,\varphi]\Bigg) \, .
\label{sugra}
\eea
Here $H_{\mu \nu \rho}=3 \partial_{[\mu} b_{\nu \rho]}$ is the curvature of the Kalb-Ramond field and $L_{m}$ is the parametrization of the matter terms in (\ref{actionDFT}). 

The action (\ref{sugra}) has an ambiguity with respect to the terms $4(\partial \phi)^2 - \frac{1}{12} H_{\mu \nu \rho} H^{\mu \nu \rho}$. These terms can be considered as part of the matter Lagrangian or as background. In this work we consider the latter in order to match with the DFT scheme, where ${\cal H}_{MN}$ and $d$ are unambiguous backgrounds. Moreover in that case we consider the set of EOMs coming from the variations of (\ref{actionDFT}) with respect to the generalized metric tensor, the dilaton and the matter fields as
\bea
\frac{\delta {\cal S}}{\delta {\cal H}^{MN}} = 0 \, , \quad \frac{\delta {\cal S}}{\delta d} =  0 \quad{\rm and} \quad \frac{\delta {\cal S}}{\delta {\Psi}} = 0 \, .\label{eoms}
\eea
Independently of the EOMs, there exist the DFT differential Bianchi identities which read \cite{FrameDFT, BianchiDFT}
\bea
\nabla_{\underline P}{\cal R} - 4 \nabla^{\overline M} {\cal R}_{\underline P \overline M} & = & 0 \, , \nn \\ \nabla_{\overline P}{\cal R} + 4 \nabla^{\underline M} {\cal R}_{\underline M \overline P } & = & 0 \, .
\label{Bianchi}
\eea

On the other hand, we know that the action must be invariant under diffeomorphism transformations. Let us take the variation of the complete action as \cite{Currents, Thermo}
\bea
\delta_\xi {\cal S}&=&\int d^{2D}X \left\{\frac{\delta \left(e^{-2d}{\cal L}\right)}{\delta d} \delta_\xi d+\frac{\delta \left(e^{-2d}{\cal L}\right)}{\delta {\cal H}^{MN}} \delta_\xi {\cal H}^{MN}+\frac{\delta \left(e^{-2d}{\cal L}\right)}{\delta \Phi} \delta_\xi \Phi\right\}\nn\\
&=&\int d^{2D}X e^{-2d} \left\{\xi^N\nabla^M G_{MN}-\xi^N\nabla^M T_{MN}+\frac{\delta {\cal L}_m}{\delta \Psi}\delta_\xi \Psi\right\}\label{diffeoinvariance}
\eea
where we identify the generalized Einstein tensor,
\bea
G_{MN} =-\frac12\eta_{MN}{\cal R}+2\left({\cal R}_{\overline M\underline N}-{\cal R}_{\underline M\overline N}\right)\label{GMNlagrangian}
\eea
and the generalized energy-momentum tensor
\bea
T_{MN} = \eta_{MN}\left({\cal L}_m-\frac12\frac{\delta {\cal L}_m}{\delta d}\right)+2\,\Big[\overline{P}_{MK} P_{NL}-\overline{P}_{NK} P_{ML}\Big]\left(\frac{\delta {\cal L}_m}{\delta {P}_{KL}}-\frac{\delta {\cal L}_m}{\delta \overline{P}_{KL}}\right)\label{TMNlagrangian}\,.
\eea
The Bianchi identities (\ref{Bianchi}) imply that $\nabla^M G_{MN}=0$ and then the first term in (\ref{diffeoinvariance}) is zero off-shell. Moreover the third term also vanishes if we apply the matter EOM. Since the complete variation is vanishing due to the invariance of the action, we conclude that 
\bea
\nabla^M T_{MN}=0
\eea
when the matter field equations are considered \cite{Park}. It is important to observe that these tensors $G_{MN}$ and $T_{MN}$ are not symmetric. Nonetheless we write down each symmetric version as
\bea
{\cal G}_{MN}=G_{M P}{{\cal H}^{P}}_N\quad {\rm and}\quad {\cal T}_{MN}=T_{MP}{{\cal H}^{P}}_N.
\eea

Finally the equations of motion (\ref{eoms}) can be cast in the single Einstein-type equation
\bea
{\cal G}_{MN}={\cal T}_{MN}\label{Einsteintypeeq}\,.
\eea
In this work we study the possibility to set a symmetric generalized energy-momentum tensor, consistent with the RHS of (\ref{Einsteintypeeq}), coming from the second moment of a generalized distribution function in a duality invariant kinetic theory. In consequence we discuss the elemental blocks that are needed to construct a double phase space and its dynamics.

\section{Double Kinetic Theory}
\label{DKT}

\subsection{The double phase space}

 Similarly to the general relativistic kinetic formalism summarized in Section 2, we define the notion of double phase space as an extension of the double space, 
 \bea
\Big\{X^{M},{\cal P}^{M} \Big\} \, ,
\eea
where for each point of the double space we consider its double tangent space whose vectors are the generalized momenta ${\cal P}^{M}$. Further the momenta ${\cal P}^M$ are $O(D,D)$ vectors. It is well known that the double space is not a (double) manifold and in consequence we demand 
\bea
\frac{\partial {\cal P}^{M}}{\partial X^{N}} = 0 \, ,
\eea
similarly to (\ref{IndepGR}).

We set a particular deformation of the  generalized diffeomorphisms
\bea
\delta_{\xi} V^{Q}(X,{\cal P}) = {\cal L}_{\xi} V^Q(X,{\cal P}) + {\cal P}^{N} \frac{\partial \xi^{M}}{\partial X^{N}} \frac{\partial V^Q(X,{\cal P})}{\partial {\cal P}^{M}} - {\cal P}^{N} \frac{\partial \xi_{N}}{\partial X_{M}} \frac{\partial V^Q(X,{\cal P})}{\partial {\cal P}^{M}}  \, ,
\label{trans}
\eea
where $\xi^{M}=\xi^{M}(X)$ and $V^Q(X,{\cal P})$ is a generic vector of the double phase space.  The expression (\ref{trans}) is trivially extended to tensors in the double phase space with different index structure. The section condition for the momentum derivatives is
\bea
\left(\frac{\partial}{\partial {\cal P}^{M}} \star\right) \left(\frac{\partial}{\partial {\cal P}_{M}} \star\right) = \frac{\partial}{\partial {\cal P}^{M}} \left(\frac{\partial}{\partial {\cal P}_{M}} \star\right) = 0 \,   
\eea
and for the mixed derivatives is
\bea
\left(\frac{\partial}{\partial X^{M}} \star\right) \left(\frac{\partial}{\partial {\cal P}_{M}} \star\right) = \frac{\partial}{\partial X^{M}} \left(\frac{\partial}{\partial {\cal P}_{M}} \star\right) = 0 \, . \,   
\eea
Imposing the previous constraints, and its simplest solution
\bea
\frac{\partial}{\partial {\cal P}_{\mu}} = 0 \,,
\eea
is enough to recover the usual phase space diffeomorphisms (\ref{ediffeos}) from (\ref{trans}) and thus ${\cal P}^{\mu}=p^{\mu}$ is the ordinary momentum.

The closure of (\ref{trans}), 
\bea
\Big[\delta_{\xi_1},\delta_{\xi_2} \Big] V^{M}(X,{\cal P}) = \delta_{\xi_{21}} V^{M}(X,{\cal P}) \,,
\eea
is given by the C-bracket
\bea
\xi^{M}_{12}(X) = \xi^{P}_{1} \frac{\partial \xi^{M}_{2}}{\partial X^{P}} - \frac12 \xi^{P}_{1} \frac{\partial \xi_{2P}}{\partial X_{M}} - (1 \leftrightarrow 2)
\eea
as in ordinary DFT (See Appendix \ref{Clos}). 

\subsection{Generalized transfer equations and conservation laws}\label{conservationlaws}

The new terms that we add to the standard generalized diffeomorphisms now force us to define the natural extension of the covariant derivative on the double phase space, namely the generalized Liouville operator. Inspecting the form of (\ref{trans}) and the transformation rule of $\Gamma_{MNP}$ it is possible to conclude that this operator must be
\bea
{D}_{M} = \nabla_{M} - \Gamma_{M N}{}^{Q} {\cal P}^{N} \frac{\partial}{\partial {\cal P}^{Q}} \, ,
\label{ext}
\eea
in order to the covariant derivative of a generalized phase space scalar transforms as a generalized phase space vector. The operator $\nabla_M$ in (\ref{ext}) is the usual DFT covariant derivative defined in (\ref{covderdft}).

Further we define the generalized distribution function $F$ as a function $F=F\big( X,{\cal P} \big)$ and, analogously to (\ref{boltzmanneq}), we propose that its evolution equation is given by the generalized Boltzmann equation that reads 
\bea
{\cal P}^{M} {\cal D}_{M} F = {\cal C}[F] \, ,
\label{BoltzmannDFT}
\eea
where ${\cal C}[F]$ is the generalized collision term and the operator ${\cal D}_M$ is
\bea
{\cal D}_{M} = D_{M} - {\cal U}_{M}\,,
\eea
with ${\cal U}_{M}= 2\, \partial_M d$. The inclusion of this last term in the generalized Boltzmann equation shows the role of the generalized dilaton when matter is taken into account. Since the integral measure is a function of the generalized dilaton, it is reasonable to expect an explicit dilaton-matter relation to appear (even within a minimally coupling scheme). In this context the generalized dilaton acts as a universal force in any element of volume of the double phase space. As we shall see below it ensures the conservation law of any moment of the Boltzmann equation for an equilibrium state.

As we have done in Section \ref{RKT}, we shall compute the transfer equations by an analogous procedure but within this generalized scheme, where the naturally invariant volume of the double phase space is
\bea
e^{-2d} d^{2D}X \,e^{-2d} d^{2D}{\cal P}=e^{-4d} d^{2D}X d^{2D}{\cal P} \, .
\eea

Let us integrate the product of a generic 1-index object $\Psi^{M}$ with the equation (\ref{BoltzmannDFT}) over all the double phase space. By assuming the existence of an equilibrium state such that $F=F_{\rm eq}$ and ${\cal C}[F_{\rm eq}]=0$ we get
\bea
\int  \Psi^{M} {\cal P}^{N} {\cal D}_{N}F_{\rm eq}\, e^{-4d} d^{2D}X d^{2D}{\cal P} = \int \Psi^{M } {\cal C}[F_{\rm eq}]\, e^{-4d} d^{2D}{\cal P} d^{2D}X =0\, .
\label{transferDFT}
\eea

We are interested in the derivation of the transfer equations coming from the expression (\ref{transferDFT}). Following (\ref{partialintegrationcoordinates}) we demand that the partial derivatives with respect to the double phase space variables, ${\cal A^M}=(X^M\,\,,{\cal P}^M)$, fulfill a divergence-type property, such that
\bea
\int d^{2D}{\cal A}\,\frac{\partial}{\partial {\cal A}^{\cal M}}\Big(e^{-2d}\,\star\Big)=0 \, .
\eea
In \cite{Thermo, Canonical} the generalized version of the Stokes' theorem for the double space was formally derived relating a divergence term integral with a boundary integral. Here we get rid of the boundary integrals by trivial boundary conditions at infinity. The final expression is independent of the integration over the coordinate volume $e^{-2d} d^{2D}X$, and it reads
\bea
&& \nabla_{N}\Big[\int \Psi^{M} {\cal P}^{N} F_{\rm eq} e^{-2d} d^{2D}{\cal P} \Big] - \int F_{\rm eq} {\cal P}^{N} {D}_{N} \Psi^{M} e^{-2d} d^{2D}{\cal P} = 0 \, .
\label{lawDFT}
\eea

If we set the arbitrary function $\Psi^{M}$ to a constant scalar, we may define, again by analogy with Section \ref{RKT}, the generalized particle current as
\bea
 {\cal N}^{M}=\int {\cal P}^{M} F_{\rm eq} e^{-2d} d^{2D}{\cal P}
\eea
with the following conservation law
\be
\nabla_{M} {\cal N}^{M} = 0 \, .
\label{c1DFT}
\ee

Considering $\Psi^{M}={\cal P}^{M}$ in (\ref{lawDFT}) leads to the generalized energy-momentum tensor 
\bea
 {\cal T}^{M N}(X)=\int {\cal P}^{M} {\cal P}^{N} F_{\rm eq} e^{-2d} d^{2D}{\cal P}
\label{TMNBoltz}
\eea
and its conservation law
\bea
\nabla_{M}{\cal T}^{M N} = 0  \, . 
\label{c2DFT}
\eea

Similarly to what occurs in GR, the present results show that the RHS of the generalized Einstein equation (\ref{Einsteintypeeq}) has the same fundamental properties than the generalized symmetric energy-momentum tensor coming from the double kinetic theory, namely it is symmetric and divergenceless. Therefore it is formally possible to replace the RHS of the Einstein-type equation (\ref{Einsteintypeeq}) with different matter contents described by the double kinetic theory, and to consistently solve the dynamics of the backgrounds and matter. 

At this point we stress that the current formulation of DFT has not a clear principle of least action for the generalized trajectories of freely falling particles with a covariant geodesic equation related to the generalized affine connection (\ref{covderdft}). In addition the generalization of the Hamiltonian function (\ref{ExplicitHamiltonianGR}) associated to the generalized Liouville operator may not coincide with the one of the plausible generalized mass-shell condition. In consequence we warn that the generalization of the Liouville vector (\ref{Liouvillevector}) may have a cumbersome relation with the present development of the double kinetic theory. Moreover the existence of a symplectic structure in the canonical formulation of the vacuum DFT has been studied in \cite{Thermo}. The inclusion of a perfect fluid in that formalism may be the natural way to define the analogous to (\ref{Hamilton}).

On the other hand it is remarkable that the generalization of the Boltzmann equation can be achieved from the invariance under generalized infinitesimal phase space diffeomorphisms, as we have shown. Further the trajectories determined by the operator ${\cal P}^MD_M$ should be consistent with a generalization of the auto-parallel curves of the DFT connection. We conjecture that future studies of the double phase space here presented may unravel other basic aspects of the double geometry.

\subsection{Applications}
\label{Applications}
The geometry of DFT can be interpreted as an $O(D,D)$ invariant extension of the Riemaniann geometry. In this sense the double geometry allows to accomodate the universal NS-NS sector of superstring theory and therefore many vacuum solutions have been explored in the last years (the incorporation of the RR-sector and sypersymmetry was also possible \cite{RR,Susy}). We point out the DFT fundamental string \cite{F1} and the DFT monopole \cite{F1, Mono}, among other solutions which have been studied in the canonical formalism of DFT \cite{Currents,Canonical}. Particularly in \cite{F1} authors considered the following split of the generalized coordinates $X^M=(t,z,y^m,\tilde{t},\tilde{z},\tilde{y}^{m})$ and the  parametrization for the generalized metric
\begin{align}
ds^2 = &\, (H-2) [dt^2 - dz^2] + \delta_{mn} dy^m dy^n + 2(H-1) [dtd\tilde{z} + d\tilde{t}dz] \nn \\ &- H[d\tilde{t}^2 - d\tilde{z}^2] + \delta^{m n} d\tilde{y}_{m} d\tilde{y}_{n} \, , \label{Gmetric2}
\end{align}
where $H$ is a harmonic function
\bea
H = 1 + \frac{h}{r^{d-4}} \, ,
\eea
$r^2=y^{m} y^{n} \delta_{m n}$ and $h$ is a constant, while the generalized dilaton is constant. Substituting (\ref{Gmetric2}) in (\ref{scalarDFT}) it is straightforward to probe that ${\cal R}=0$. Moreover, in this geometry ${\cal G}_{MN}=0$ \footnote{See appendix A of \cite{F1} for the explicit computation.} and thus, also from the perspective of the generalized Einstein equation, this is a vacuum solution. A suitable deformation of the construction (\ref{Gmetric2}) is mandatory in order to couple matter. Analogously, the DFT monopole \cite{Mono} also satisfies ${\cal G}_{MN}=0$.

According to the RHS of the generalized Einstein equation, the simplest matter to introduce from a double kinetic theory perspective is a perfect fluid. 
The generalized energy-momentum tensor can be computed either integrating the generalized distibution function of the system (\ref{TMNBoltz}) or using the variational principle (\ref{diffeoinvariance}) with a consistent matter Lagrangian. Since the results of this work are not enough to perform the former, we are force to work with a field equivalent to the perfect fluid.

A first proposal in this line is to minimally couple an $O(D,D)$ invariant scalar field to the background content of DFT considering a matter Lagrangian as follows
\bea
{\cal L}_{m}[{\cal H},\Phi] = - \frac12 {\cal H}^{M N} \nabla_{M} \Phi \nabla_{N} \Phi \, .
\label{Scalarlm}
\eea
The equation of motion of the generalized scalar field is the equivalent of the Klein-Gordon equation namely
\bea
{\cal H}^{M N} \nabla_{M} \nabla_{N} \Phi = 0 \, .
\label{KGDFT}
\eea
Moreover, let us observe that ${\cal L}_{m}$ and the generalized Klein-Gordon equation reduce to its standard versions using the parametrization (\ref{Gmetric}) and the strong constraints. The generalized symmetric energy-momentum tensor reads
\bea
{\cal T}_{M N} = - 4\, \overline{P}_{K(M}\,P_{N)L}\,\nabla^{K} \Phi \nabla^{L}\Phi - \frac12 {\cal H}_{M N} {\cal H}^{R Q} \nabla_{R}\Phi \nabla_{Q}\Phi \, .
\label{ScalarTMN}
\eea
From the double kinetic side we have an equilibrium system such that,
\bea
\int ({\cal P}^{\un M} {\cal P}^{\un N} + {\cal P}^{\ov M} {\cal P}^{\ov N} ) F_{\rm eq} e^{-2d} d^{2D}{\cal P} & = & - \frac12 {\cal H}_{M N} {\cal H}^{R Q} \nabla_{R}\Phi \nabla_{Q}\Phi \,,\label{feqovun}\\
\int ({\cal P}^{\ov M} {\cal P}^{\un N} + {\cal P}^{\un M} {\cal P}^{\ov N}) F_{\rm eq} e^{-2d} d^{2D}{\cal P} & = & - 2 \nabla_{\ov M} \Phi \nabla_{\un N}\Phi - 2 \nabla_{\un M}\Phi \nabla_{\ov N}\Phi \, .\label{feqmixed}
\eea

Using $\nabla_{\ov M}( \nabla_{\un N}\Phi) = \nabla_{\un N}( \nabla_{\ov M}\Phi)$ and (\ref{KGDFT}) on (\ref{ScalarTMN}) it is straightforward to show that
\bea
\nabla^{M}{\cal T}_{M N} = 0 \, ,
\eea
in agreement with the results of the double phase space formulation in Section \ref{conservationlaws}.

The inclusion of (self-)interacting terms can be perform by adding a suitable potential to the matter Lagrangian as
\bea
{\cal L}_{m}[{\cal H},\Phi] = - \frac12 {\cal H}^{M N} \nabla_{M} \Phi \nabla_{N} \Phi - V[\Phi] \,. 
\eea
Now the energy-momentum tensor contains extra terms that do not affect its conservation law. Particularly it reads  
\bea
{\cal T}_{M N} = - 4\, \overline{P}_{K(M}\,P_{N)L}\,\nabla^{K} \Phi \nabla^{L}\Phi - \frac12 {\cal H}_{M N} {\cal H}^{R Q} \nabla_{R}\Phi \nabla_{Q}\Phi - {\cal H}_{M N} V[\Phi]\,.
\label{TMNfull}
\eea
In this case we observe that (\ref{feqmixed}) remains unchanged but the last term of (\ref{TMNfull}) must be added to the RHS of (\ref{feqovun}). Equations like (\ref{TMNfull}) were already used to analyze aspects of duality covariant cosmology \cite{completionf}.

In GR the equivalence between the energy-momentum tensors of a scalar field and the one of a perfect fluid is regarded as a formal correspondence and it is based on the identification of the energy density and pressure of the fluid with a expression involving the derivatives of the scalar field and its potential \cite{scalarfluidgr}. In DFT this procedure requires a deeper understanding of the intensive properties of the fluid in the double geometry and the proposal in this work is a first step into it. We leave this issue and the study of the appropriate deformation of the vacuum DFT solutions coupled to a perfect fluid for future work.

\section{Outlook}
\label{Outlook}
We present a model of kinetic theory in the context of DFT. We define a double phase space where tensors depend both on the generalized coordinates $X^M$ and the generalized momentum ${\cal P}^{M}$. Generalized diffeomorphisms on the phase space are consistently defined and the covariant derivative is replaced by a generalized Liouville operator, as it happens in ordinary general relativistic kinetic theory. The closure of the transformations is given by the C-bracket. The previous formalism allows us to introduce the analogue of the Boltzmann equation for a generalized distribution function, which describes the evolution of the number of particles in a volume element of the double phase space. It turns out that the generalized Boltzmann equation acquires an extra term due to the role of the generalized dilaton acting as a universal force in the double geometry.

We define the generalized current and energy-momentum tensor as the first and second moment of the generalized distribution function as usual. Then we derive its conservation laws and we show that both are conserved. In particular we point out that the generalized energy-momentum tensor from the double kinetic side has the same fundamental properties, it is symmetric and divergenceless, than the RHS of the generalized Einstein-type equation (\ref{Einsteintypeeq}). As an application we discuss how to couple a perfect fluid to the background content of DFT from a variational principle using a generalized scalar field. We derive a generalization of the Klein-Gordon equation which must be used to prove that the generalized symmetric energy-momentum tensor is in fact divergenceless in agreement with the double kinetic theory. The results of this work open the door to a large number of questions and future directions. We elaborate on some important points:

\begin{enumerate}[label=(\roman*)]

\item \textbf{Double Cosmology}

The construction of a low energy effective string cosmology based on DFT is a promising area of work \cite{completionf, Cosme}. One interesting aspect of this is that the inclusion of the dual coordinates $\tilde{x}$ provides that the cosmological singularities of a homogeneous and isotropic universe may disappear thanks to the T-duality symmetry. An apparent big bang singularity in the ordinary supergravity framework is (T-)dual to an expanding universe in the dual dimensions when the section condition is suitably applied. The present work could be an intermediate step in finding a manifestly T-duality invariant energy-momentum tensor for a perfect fluid in the double-space which could give a description of the geometry and the matter in double cosmology \cite{Point}.

\item \textbf{Generalized distribution functions and equilibrium}

The equilibrium states and their properties are very well-known in GR \cite{Cercignani}, particularly the equilibrium 1pdf as a function of the momentum and different lagrange multipliers. The generalization of the latter to the double space together with a suitable generalized mass-shell condition could be a great step in the description of matter in DFT. It would allow us to explicitly evaluate the generalized particle current, the generalized energy-momentum tensor and its conservation laws.

\item \textbf{Collisions in double space}

The conservation laws of the ordinary current $N^{\mu}$ and energy-momentum tensor $T^{\mu \nu}$ in GR still holds out of the equilibrium state. The key point of this statement is that these conservation laws come from the transfer equation considering a collisional invariant quantity, the momentum $p^{\mu}$ for the procedure of taking moments. It is possible that a T-duality invariant treatment of collisions in the double space could be captured using the framework presented here and we expect a generalized notion of collision-invariants related to the generalized momentum.

\item \textbf{H-theorem and thermodynamics}

The transfer equation,
\bea
\int \Psi^{\nu}(p^{\mu} \frac{\partial f}{\partial x^{\mu}} - \frac{\partial f}{\partial p^{\sigma}} \Gamma^{\sigma}_{\mu \rho} p^{\mu} p^{\rho})gd^dpd^dx = \int \Psi^{\nu} C[f] g d^dp d^dx \, ,
\label{transferend}
\eea
allows to define an entropy current $S^{\mu}$ considering
\bea
\Psi^{\mu} \propto - \ln \left(\frac{fh^3}{g_s}\right)
\eea
where $h$ is the Planck constant and $g_s$ the degeneracy factor. The previous statement is known as the H-theorem \cite{Cercignani}. In this context the conservation law of the entropy current,
\bea
\nabla_{\mu} S^{\mu} \geq 0
\eea
is understood as the second law of thermodynamics. Finding a T-duality generalization of the previous results through the generalized transfer equation (\ref{transferDFT}) could be an interesting direction to continue the present work. We expect that the previous treatment allows to establish equivalences between our model and \cite{Thermo}.

\end{enumerate}

\subsection*{Acknowledgements}
We thank D. Marqu\'es for many enlightening discussions and useful comments on the draft. EL would like to thank CONICET for supporting his work. NMG thanks to Departamento de Física, Facultad de Ciencias Exactas y Naturales, Universidad de Buenos Aires for support.

\appendix

\section{Conventions}
\label{Appendix}
In this appendix we introduce the notation used throughout the paper.

The Christoffel connection is
\bea
\Gamma_{\mu \nu}^{\rho} = \frac12 g^{\rho \sigma} (\partial_{\mu} g_{\nu \sigma} + \partial_{\nu} g_{\mu \sigma} - \partial_{\sigma} g_{\mu \nu})
\eea
while its transformation rule under infinitesimal diffeomorphisms is 
\bea
\delta_{\xi} \Gamma_{\mu \nu}^{\rho} = \delta_{\xi}^{cov} \Gamma_{\mu \nu}^{\rho} + \partial_{\mu \nu} \xi^{\rho} 
\eea
where $\delta_{\xi}^{cov}$ is the transformation rule of a (2,1) tensor.

The covariant derivative of a generic (1,1) tensor is given by
\bea
\nabla_{\rho} v_{\mu}{}^{\nu} = \partial_{\rho} v_{\mu}{}^{\nu} - \Gamma_{\rho \mu}^{\sigma} v_{\sigma}{}^{\nu} + \Gamma_{\rho \sigma}^{\nu} v_{\mu}{}^{\sigma} \, .
\eea

The trace of the connection is 
\bea
\Gamma_{\mu \nu}^{\nu}=\partial_{\mu}(\ln(\sqrt{g})) \, .
\label{tracec}
\eea

The momentum coordinates satisfies,
\bea
\frac{\partial p^{\mu}}{\partial p^{\nu}} = \delta^{\mu}_{\nu} \, . 
\eea
We use the following index convention throughout the work,
\bea
W_{(\mu \nu)} = \frac12 W_{\mu \nu} + \frac12 W_{\nu \mu} \, , \quad 
W_{[\mu \nu]} = \frac12 W_{\mu \nu} - \frac12 W_{\nu \mu} \, ,
\eea
with $W_{\mu \nu}$ arbitrary.
\section{Closure}
\label{Clos}
We show the closure for a generic vector of the double phase space,
\bea
\Big[\delta_{\xi_{1}},\delta_{\xi_{2}} \Big]V_{M} &=& \delta_{\xi_{1}} \Big(\xi_{2}^{N} \partial_N V_M + (\partial_M \xi_{2}^N - \partial^N \xi_{2M}) V_N \nn\\
&&+ \omega (\partial_{N} \xi_{2}^{N})V_{M} + \delta_{\xi_{2}}^{(p)}V_{M}  \Big) - (1\leftrightarrow2) \, .
\label{closure1}
\eea
The extension to tensors of the double phase space is straightforward. In (\ref{closure1}) we introduce the following notation,
\bea
\delta_{\xi_{2}}^{(p)}V_{M} = {\cal P}^{Q} (\partial_{Q} \xi_{2}^{R}) \frac{\partial V_{M}}{\partial {\cal P}^{R}}  
\eea
where $\partial_{M}=\frac{\partial}{\partial X^M}$. We want to show that the previous expression is equivalent to $\delta_{\xi_{21}} V_{M}$ with 
\bea
\xi^{M}_{12}(X) = \xi^{P}_{1} \partial_{P} \xi^{M}_{2} - \frac12 \xi^{P}_{1} \partial^{M} \xi_{2P} - (1 \leftrightarrow 2)  \, . 
\eea
Since the standard generalized diffeomorphisms close with the C-bracket, we just need to show that the extra terms in (\ref{closure1}) are
\bea
&& {\cal P}^{N} \partial_{N} (\xi^{P}_{2} \partial_{P} \xi_{1 Q} - \frac12 \xi^{P}_{2} \partial_{Q} \xi_{1P}) \frac{\partial V_M}{\partial {\cal P}_{Q}} - (1 \leftrightarrow 2) \, .
\label{extraclos}
\eea
Let us note that the last term of the previous expression is trivially null using the strong constraint, and therefore we need to recover only the first one.

The extra terms in (\ref{closure1}) are
\bea
&& \Big( \xi_{2}^{N} \partial_N (\delta^{(p)}_{1} V_M) + (\partial_M \xi_{2}^N - \partial^N \xi_{2 M}) (\delta^{(p)}_{1} V_N)  + \omega (\partial_{N} \xi_{2}^{N}) (\delta^{(p)}_{1} V_{M}) \nn \\ && + (\delta_{1}{\cal P}^{N}) \partial_{N} \xi_{2}^{Q} \frac{\partial V_M}{\partial {\cal P}^{Q}} + {\cal P}^{N} \partial_{N} \xi_{2}^{Q} \frac{\partial}{\partial {\cal P}^{Q}}(\delta_{1} V_{M}) \Big) - (1\leftrightarrow2) \, .\label{b6}
\eea
The first term of the second line of (\ref{b6}) is zero using the transformation of ${\cal P}$ given by (\ref{trans}) and the strong constraint. The remaining terms are
\bea
= && \Big( {\cal P}^{Q} \xi_{2}^{N} \partial_N ( \partial_{Q} \xi_{1}^{R}) \frac{\partial V_{M}}{\partial {\cal P}^{R}} + {\cal P}^{Q} \xi_{2}^{N}  \partial_{Q} \xi_{1}^{R} \partial_N( \frac{\partial V_{M}}{\partial {\cal P}^{R}}) \nn \\ && + (\partial_M \xi_{2}^N - \partial^N \xi_{2 M}) ( {\cal P}^{Q} \partial_{Q} \xi_{1}^{R} \frac{\partial V_{N}}{\partial {\cal P}^{R}}) + \omega (\partial_{N} \xi_{2}^{N}) ( {\cal P}^{Q} \partial_{Q} \xi_{1}^{R} \frac{\partial V_{M}}{\partial {\cal P}^{R}}) \nn \\ && + {\cal P}^{N} \partial_{N} \xi_{2}^{Q} \frac{\partial}{\partial {\cal P}^{Q}}(\xi_{1}^{R} \partial_R V_M + (\partial_M \xi_{1}^R - \partial^R \xi_{1 M}) V_R  + \omega (\partial_{R} \xi_{1}^{R})V_{M}) \nn \\ && + {\cal P}^{N} \partial_{N} \xi_{2}^{Q} \partial_{Q} \xi_{1}^{R} \frac{\partial V_{M}}{\partial {\cal P}^{R}} \Big) - (1\leftrightarrow2) \nn \, ,
\eea
where we have used that $\frac{\partial P_M}{\partial P_{N}} = \delta_{M}^{N}$. Up to this point is easy to note that the closure does not depend on the weight factor $\omega$. The terms with two derivatives acting on $V_{M}$ also simplify and we have,
\bea
= && \Big( {\cal P}^{N} \xi_{2}^{Q} \partial_Q ( \partial_{N} \xi_{1}^{R}) \frac{\partial V_{M}}{\partial {\cal P}^{R}} + {\cal P}^{N} \partial_{N} \xi_{2}^{Q} \partial_{Q} \xi_{1}^{R} \frac{\partial V_{M}}{\partial {\cal P}^{R}} \Big) - (1\leftrightarrow2) \nn \\ = && \Big( {\cal P}^{N} \partial_{N} (\xi^{Q}_{2} \partial_{Q} \xi_{1 R} ) \frac{\partial V_M}{\partial {\cal P}_{R}} \Big) - (1 \leftrightarrow 2) \, ,
\eea
which matches with (\ref{extraclos}).

\end{document}